# Wall-plug (AC) power consumption of a very high energy e+/e- storage ring collider

Marc Ross, SLAC

August 3, 2013

Estimated AC power consumption ranges from ~280 MW (1) to 416 MW (2) for a 350 GeV $E_{cm}$ 80 km circumference colliding beam storage ring complex with parameters given in (1). The difference between the two estimates is from differing assumptions concerning heat removal, cryo-plant efficiency, klystron operation etc. The purpose of this note is to list and explain these.

RF and Utility power consumption

1) 'Saturated' klystron operation is very unusual in storage rings. At LEP this was done only for the last year of operation in an attempt to capture as much luminosity at the highest achievable energy and is not a reasonable approach to take for a new machine. During that last LEP-year, any perturbation generated a 'beam-trip' (3). Reducing the klystron drive provides the flexibility needed for feedback operation but results in a need for more klystrons and reduced klystron efficiency. For an overhead of 7%, the efficiency is reduced to 55% from 65% (32 MW). (4)
2) Cryo – plant power required at JLab for 1.9 degrees K is 1100 W per watt dissipated at low temperature (5). This is 20% worse than the assumed value in (1) (7 MW). (CERN systems have somewhat better performance.)
3) Following the Swiss Alps tunnel fires CERN began to consider actively controlled 'transversal' tunnel ventilation systems (6). Such systems are much safer as they allow segmented control of tunnel air flow but are more expensive and require more extensive ventilation equipment. The much simpler LHC ventilation system was allowed to remain and was 'grandfathered'. If it were upgraded, the power consumption would be more than two times what is used today, rising to 14 MW. The larger 80 km ring would require three times this, 42 MW, to ventilate.
4) The electrical power required to remove heat through primary and secondary water cooling system loops in typical accelerator installations is between 5 and 10%; i.e. 5 watts is needed to power machinery to remove 100 watts of dissipated power. This depends on environmental conditions, the required water volume, and the number of stacked loops, (cooling tower is the primary loop and low-conductivity water is a secondary loop. A third loop is often required if the component cooling water can be activated.) RF system power is dissipated in the power supply, klystron collector and synchrotron radiation. For a sub-surface ring, the power will be dissipated below-grade and will require water cooling between 5 and 10% (10 to 20 MW) (7). For large, distributed systems, required pumping power will be larger. For magnet systems we should assume 10% is required.
5) Electrical network losses, which may be substantial for a distributed complex, will be about 5% (8). These losses are not included here.

Table 1: Estimated AC power consumption for a 350 GeV E_cm 80 km circumference colliding beam storage ring complex

| Power Consumption (MW) | TLEP 175 (1) Tables 2 and 3 | Corrections (1 – 4) to TLEP 175 | Corrected TLEP-175 |
|---|---|---|---|
| SRF power to beam – collider | 100 | | 100 |
| Klystron system – collider | 185 | 32 | 217 |
| Cryogenics | 34 | 7 | 41 |
| Top-up ring (RF) | 5 | 0 | 5 |
| Water cooling (SR / Klystron) | 5 | 15 | 20 |
| Ventilation | 21 | 21 | 42 |
| Magnet systems | 14 | (See below) | 14 |
| General services | 20 | | 20 |
| **Total** | **284** | **75** | **359** |

Two collider rings or one?

Are there two collider rings or just one? Is a single overall design scheme adequate for both E_cm=350 and E_cm = 90, at high luminosity? For the highest energy operation two rings are required because of the energy-loss 'saw tooth' and for low energy high current operation two are required because of the large current and large number of parasitic crossings. Both PEP-II and KEK-B designs, with current similar to 90 GeV E_cm parameters, are based on two rings. Above about 100 mA (9% of nominal 90 GeV E_cm) multi-cell RF cavities are not used because of trapped higher-order-modes (9). Both PEP-II and KEK-B use heavily damped single-cell RF cavities with a packing-factor 5 to 10 times worse than multi-cell cavities. The RF effective length then becomes 3 to 6 km and the cost of the SRF system would scale accordingly.

Magnet Power consumption

Estimated collider ring magnet power consumption is the single largest discrepancy between the two estimates cited above. For E_cm 350 GeV, there is no valid collider ring lattice design. We can assume the number of cells in the lattice to be substantially increased, compared to LEP or LHeC, in order to achieve the needed momentum acceptance of 2.5%. For modern synchrotron light ring optics this can be about a factor of two to five. This factor, applied to two instead of one ring, gives 12 times (or 30 times) as many magnets as listed in the LHeC design giving a ring power consumption of 42 MW, 3 times larger than the listed value of 14 MW. This is the largest difference between the two estimates.

Table 2: Magnet system and injector system power consumption

| Collider Ring magnet consumption (MW) | TLEP 175 (1) Table 3 (one ring) | Two rings, # optics cells ~ 2x LEP |
|---|---|---|
| Magnet systems | 14 | 42 |
| Power required for water cooling (10%) |  | 5 |
| Magnet total | 14 | 47 |
| Total including Table 1 | 284 | 392 |
| Injector complex (10) |  | 16 |
| **Total Ring-Collider power consumption** | **284** | **408** |

Total CERN energy consumption in 2012 was 1.35 TWh, 15% of which was for the campus. For 5000 hours operation (typical, not verified) this is an average power consumption of 230 MW.

References

(1) Mike Koratzinos, IPAC13, http://accelconf.web.cern.ch/AccelConf/IPAC2013/papers/tupme040.pdf
(2) Marc Ross, 'Higgs Quo Vadis', https://indico.cern.ch/contributionDisplay.py?sessionId=5&contribId=60&confId=202554
(3) Steve Myers, Personal communication, 8 July, 2013
(4) Chris Lingwood, ESS Workshop Uppsala 2011, slide 33
(5) Tom Powers, ICFA Beam Dynamics Newsletter 60, 2013, http://icfa-usa.jlab.org/archive/newsletter/icfa_bd_nl_60.pdf
(6) Fabio Corsanego, http://indico.cern.ch/getFile.py/access?contribId=29&sessionId=1&resId=1&materialId=slides&confId=165201
(7) Richard Neal, 'The Stanford Two Mile Accelerator', pg 935 (Ch. 24).
(8) Bernard Jeanneret and Philippe Lebrun, CLIC workshop, January 2013, slide 6. https://indico.cern.ch/getFile.py/access?contribId=68&sessionId=24&resId=1&materialId=slides&confId=204269
(9) Andy Butterworth, TLEP workshop, January 2013, slide 30
(10) SSC injector power consumption estimates (linac + LEB + MEB)